**Intensity tuning of the edge states in the imperfect topological waveguides
based on the photonic crystals with the $C_3$ point group symmetry**

J. Hajivandi[1,*], H. Pakarzadeh[2] and H. Kurt[1]

[1]Nanophotonics Research Laboratory, Department of Electrical and Electronics Engineering, TOBB University of Economics and Technology, Ankara 06560, Turkey
[2]Department of Physics, Shiraz University of Technology, Shiraz, Iran
[*]*jamileh@etu.edu.tr*

**Abstract:** In this paper, we explore the topological behavior of a two-dimensional honeycomb photonic crystal (PC) based on the presence of double Dirac-cone connected the orbitals $p$ and $d$, due to the $C_6$ point group symmetry of the hexagonal PCs. Removing the four-fold degeneracies between the bands at the Dirac point can be achieved by introducing three small dielectric rods near the bigger ones to realize the perturbed PCs with the $C_3$ point group symmetry with different topological features. By proposing the unique structure involving two PCs with different topological effects, one may study the one-way light distribution along the local boundary in spite of the defects, cavities, and disorders. Moreover, we investigate the variation of the transmitted intensity values under different defect conditions and realize that the size, location and material type affect the transmitted light. In the other words, tunable intensity of the edge states can be achieved through adjusting the defects such that by increasing the radius of rods, the intensity of the edge states can be enhanced or by increasing their distance from the unit cell center, the intensity will be decreased. For example, by increasing the radius of one of the rods, the intensity of the edge states is increased up to the 130%. Additionally, by moving the position of the mentioned rod along the direction, x, the intensity of the edge states will be decreased to 82.08%. Finally, topological rhombic resonator enables unidirectional filtering of guided mode. The fact that different light manipulation scenarios can be realized provides a unique aspect for topological photonic insulators.

1. Introduction

   Photonic crystals (PCs) and their applications for various purposes like optical cavities, waveguides and mirrors based on their dispersion relations, have been investigated for a long time. Discovering of the topological behaviour of the materials in condensed matter physics provides an idea to explore the properties of the PC band structures for employing them as photonic topological insulators [1–3]. Although the spontaneous breakdown of symmetries cannot justify the band dispersion theory, it can completely express the innovative topological properties in materials [3-7]. The topological behaviour of PCs can be realized through breaking the time-reversal (TR) symmetry or without breaking the TR. For instance, the first approach can be applied in the coupled helical mechanism where the magnetic excitation is required while the second approach may be used in the spin-polarized unidirectional distribution of the surface photons and the bianisotropic photonic metacrystals [8–24]. For example, Khanikaev et. al. have studied the outstanding and valuable characterization of time-reversal-invariant topological insulators protecting one-way spin polarized transport along the interface as the important part of their research published in Nature Materials in 2012 [25]. Using the photonic metacrystal in the presence of the Magneto-electric coupling due to the electromagnetic duality between the electric and magnetic fields, they found the symmetry protected topological transitions [25]. It can be demonstrated that there is an inherent pseudo TR symmetry in the PCs preserving the $C_6$ point group symmetry condenses the Kramers doublet as well the TR symmetry supported by Maxwell equations. The topological phase of the PC can be evaluated by calculating the Berry curvatures of the edge modes along with the bulk photonic bands. So the PCs preserving the $C_6$ point group symmetry with simple structure for fabrication, without needing any external field, support the photonic topological edge states [26]. Moreover, the Quantum-Valley-Hall effect has been demonstrated in an All-Silicon triangular PC with broken inversion symmetry. Indeed, TE polarization has been used for raising the valley degree of freedom. So a manageable bandgap splitting opposite topological phases and transport may be realized by breaking the inversion symmetry of the PC [26]. The topological behaviour of the PCs can be applied for various purposes such as light confinement in photonic devices. Indeed, putting two PCs with the same band gaps but different topological behaviors together, results in appearing the photonic edge modes propagating along the interface [27–35]. Hence, the topological PCs made of dielectric materials (instead of metal or magnetic materials) can be used for optical applications, because they exhibit robust and superior edge states [36-39].

In our recent works, we studied the behavior of the transmitted light in the vicinity of the various types of defects and cavities at the topological interfaces. Furthermore, the directional surface mode exiting from the topological material as well as the creation of Fano resonance mode and its preserving in the photonic topological configurations have been studies [40-43].

In this paper, we investigate the topological characterization of a two-dimensional (2D) hexagonal PC preserving the TR symmetry associated with the $C_6$ point group symmetry. Introducing three smaller rods near the main ones, the symmetry of the structure can be decreased to $C_3$ which leads to opening the Dirac cone at the central point of the Brillouin zone, $k = 0$. By studying the behavior of the PC before and after the Dirac cone, a topological phase transition can be realized which is the main key for studying the topological properties of the configuration. By designing a topological structure composed of the perturbed PCs, the one-way scattering-free propagation of the edge states at the topological interface is realized. Also, the intensity tuning of the edge states is performed by modifying the size and position of the rods in the defect topological structures. Therefore, by increasing the radius of the defect rod, the intensity will be enhanced or by increasing the moving the defect rod from the initial position, the intensity of the edge states will be decreased. Besides, by introducing the Al blocks in front of the source and by increasing its width, the intensity will be decreased. Additionally, the scatter free transport of the edge modes around the sharp corners like rhombic resonator will be explored.

## 2. All-dielectric two-dimensional modified honeycomb lattices

We start with the ordinary 2D honeycomb PC that is constructed with two lattice vectors $\vec{a}_1 = (a, 0)$ and $\vec{a}_2 = \left(\frac{a}{2}, \frac{a\sqrt{3}}{2}\right)$ where $a$ is the lattice constant. Each unit cell is composed of six dielectric rods with permittivity of $\varepsilon = 12$ and radius $r$ where the center is located at a distance of $R$ from the center of unit cell as shown in Fig. 1(a). By tuning the geometrical parameters as $R = a/3$ and $r = R/3$, the structure exhibits the $C_6$ point group symmetry which leads to appearing the accidental double Dirac cone at the Brillouin zone center, $\Gamma$ point. Many approaches in terms of increasing/decreasing the radii of the rods or changing the locations have been proposed in the literature to remove the degeneracy at the Dirac point [32,44-45]. In the present work, we propose a new approach to modify the band diagrams and investigate topological photonic insulators. By introducing three small rods near the bigger ones, we can break the inversion symmetry. This perturbation leads to reducing the point group symmetry from $C_6$ to $C_3$ as shown in the PCs type-A and type in Fig. 1(b). The small rods are chosen with the electrical permittivity $\varepsilon=12$ and the radii $r_1 = R/7$ with the distance of $R_1 = R - r - r_1$ from the unit center.

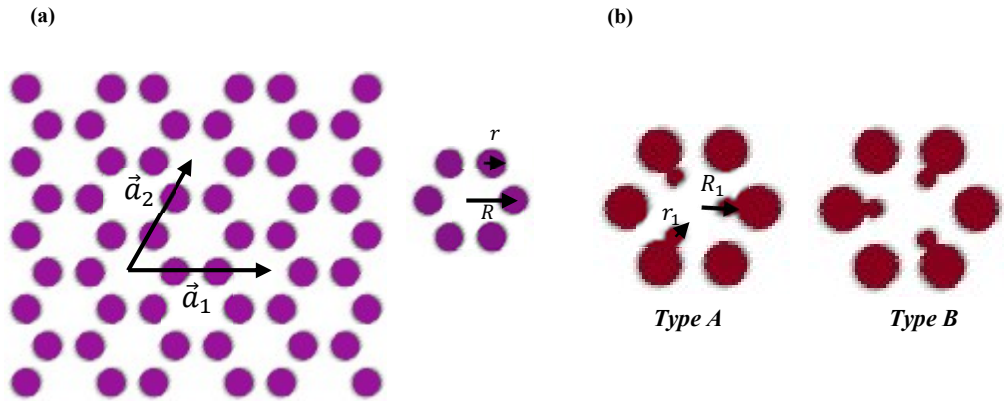

Fig. 1. (a) The schematic geometry of the 2D honeycomb PC. The vectors $\vec{a}_1$ and $\vec{a}_2$ are the basis lattice vectors. The radius of each rod is $r$ and distance from the unit cell center is $R$. (b) Two modified PCs named as type $A$ and $B$, six main rods with radii of $r$ are embedded at a distance of $R$ from the center of the unit cell, while three small rods with radii of $r_1$ are embedded at a distance of $R_1 = R - r - r_1$.

The dispersion behavior of the 2D honeycomb PC is shown in Fig. 2(a) for the parameters, $R = a/3$ and $r = R/3$. As it is seen, at the double Dirac cone, four degenerate bands cross each other and the structure exhibits the $C_6$ point group symmetry which leads to appearing the accidental double Dirac cone at the Brillouin zone center, $\Gamma$ point. This is one of the simplest lattices used to studying the topological transitions without breaking the TR symmetry. Here, we study the band structures for the TM mode ($E_z, H_x, H_y \neq 0$), using the software package *MIT Photonics Bands*

(MPB) [46]. Fig. 2 (b) indicates the band diagrams of the two modified honeycomb PCs whose unit cells are seen in Fig. 1(b). As it is seen, the photonic band gap is appeared owing to the lifting of the four-fold degeneracy which is indicated by points b and p in Fig. 2(b). Yet, there is still a two-fold degeneracy. The electric field profiles of the modified honeycomb PCs are designated as symmetric and anti-symmetric and also they are converted to each other along the green vectors as given in Fig. 2(c).

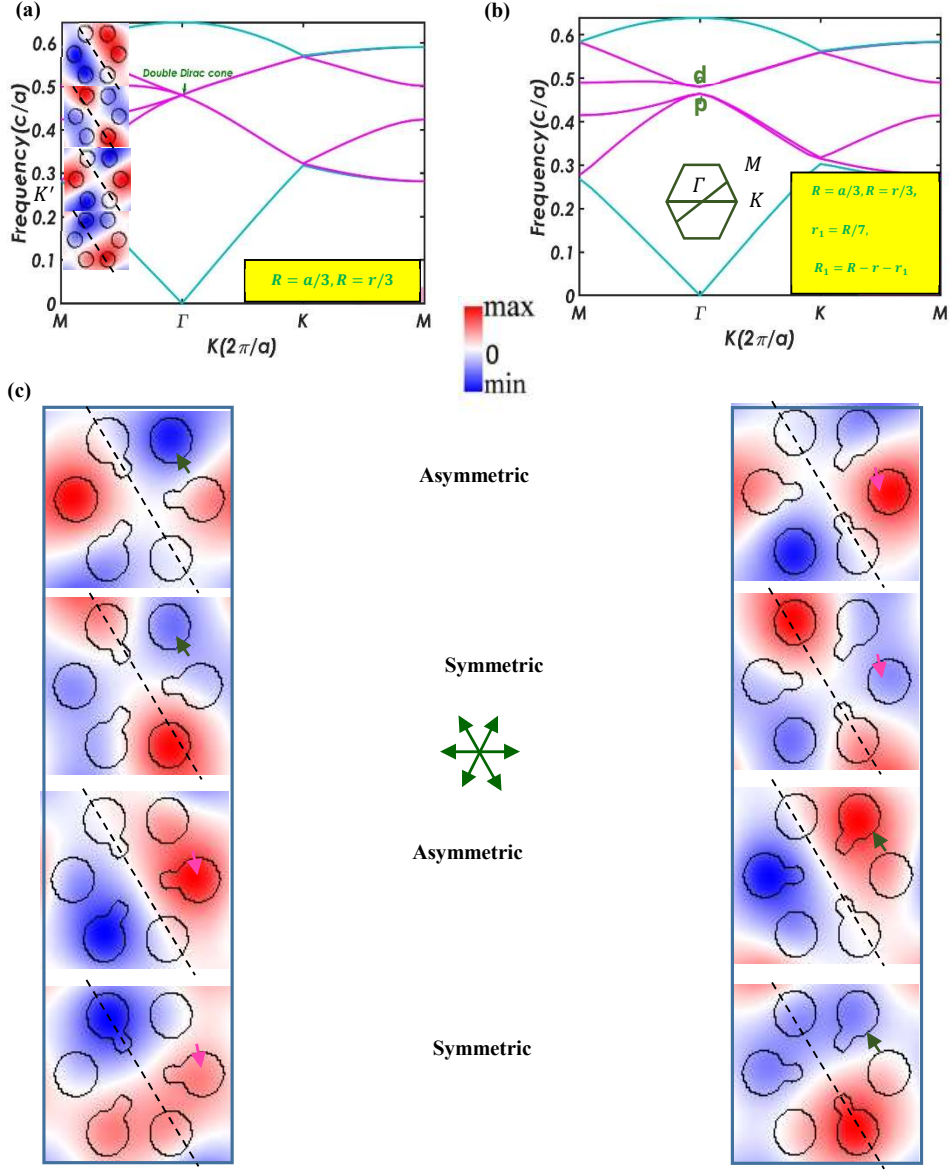

Fig. 2. (a) The band diagram of two modified HPCs, types $A$ and $B$, by introducing small rods with radii of $r_1 = R/7$ at the distance of $R_1 = R - r - r_1$. (b) The dispersion diagram of the HPC for the parameters $R = a/3$ and $r = R/3$. The inset is the Brillouin zone of the lattice. The $E_z$ profiles of the degenerated bands are shown as inset. (c) The electric field profiles of the two-fold degenerate bands for the modified honeycomb PCs type $A$ and $B$ with the phase difference, $\pi$.

In this paper, excitation of the topological edge mode protecting the pseudospin dependent propagation, due to lifting the double Dirac cone is desired. Decreasing the point group symmetry of the PC from the $C_6$ to $C_3$ leads to lifting the double Dirac cone, then inducing the topological edge states transport. Indeed, we design a microscale PC demonstrating the topological states in the absence of any outside magnetic fluidity. Then, the accidental four-fold degenerated double Dirac cone arising from the junction of dipole modes ($p$) and quadrupole modes ($d$) have been observed at the central point of the Brillouin zone, $\Gamma$, associated with the unavoidable three-fold degenerated Dirac cone at the $K(K')$ point. By breaking the inversion symmetry, the double Dirac cone is opened, separating the dipole and quadrupole modes due to decreasing of the point group symmetry from $C_6$ to $C_3$. Generally, evaluating the Chern numbers in $C_6$ point group symmetry, is desired using the $k.p$ perturbation model. Indeed, we can abbreviate the Chern number as $C_\pm = \pm(sign(B) + sign(M))/2$, by calculating the effective Hamiltonian nearby the point $\Gamma$. Where $B$ is mainly negative resulting from the diagonal components of the second order perturbation and $M = \frac{1}{2}(E_2 - E_1)$ where $E_1$ and $E_1$ are the Eigen modes of the orbitals $d$ and $p$ respectively. Then, $M > 0$ if the orbital $d$ becomes higher than the orbital $p$ in the band diagram of the PC, leads to $C_\pm = 0$. Against, if the orbital $p$ becomes higher than orbital $d$, we would have $M < 0$, results in $C_\pm = \pm 1$. Therefore, the topological phase transition from trivial PC with $C_\pm = 0$ to topological (non-trivial) one with $C_\pm = \pm 1$ will be induced. In our work, Although the orbital d is higher than orbital p in the electric field diagrams of both types A and B, it seems that there is the phase difference between the field profiles which yields in topological phase transition. For more details, compare the right and left profiles, $E_z$ in Fig. 2(c). The field profiles of the PC, type B (right) can be extracted from the field profiles of the PC, type A (left), by the phase difference $\pi$. Then we can claim that a topological phase transition has been supplemented and we can realize the robust edge mode at the domain interface of the PC. Introducing the small rods with radii, $\frac{R}{7}$, near the big ones, then reducing the $C_6$ point group symmetry to $C_3$, doesn't have noticeable influence on opening the band gap at $K(K')$ points. Then, we ignore more studying on the opposite berry curvatures at the $K(K')$ points. For more details about this topic refer to the ref. [26]. We have also investigated that the proposed design of HPC may be applied for other values of radii other than $r_1 = R/7$. Fig. 3 shows the topological band gap for various radii $r_1$. In all of these figures, the parameters $R = a/3$ and $r = R/3$ are fixed and the distance between the center of the small rods from the center of the unit cell are adjusted to $R_1 = R - r - r_1$. For example, for radius of $r_1 = R$, the orbitals $d$ and $s$ become degenerate and for radii of $r_1 = R/2$ and $r_1 = R/3$, the orbital $s$ embedded between orbitals $p$ and $d$. Also, for radii of $r_1 = R/4$, $r_1 = R/5$ and $r_1 = R/6$, there is a clear gap between orbitals $p$ and $d$ and for radii of $r_1 = R/8$, $r_1 = R/9$ and $r_1 = R/10$, there is a small band gap.

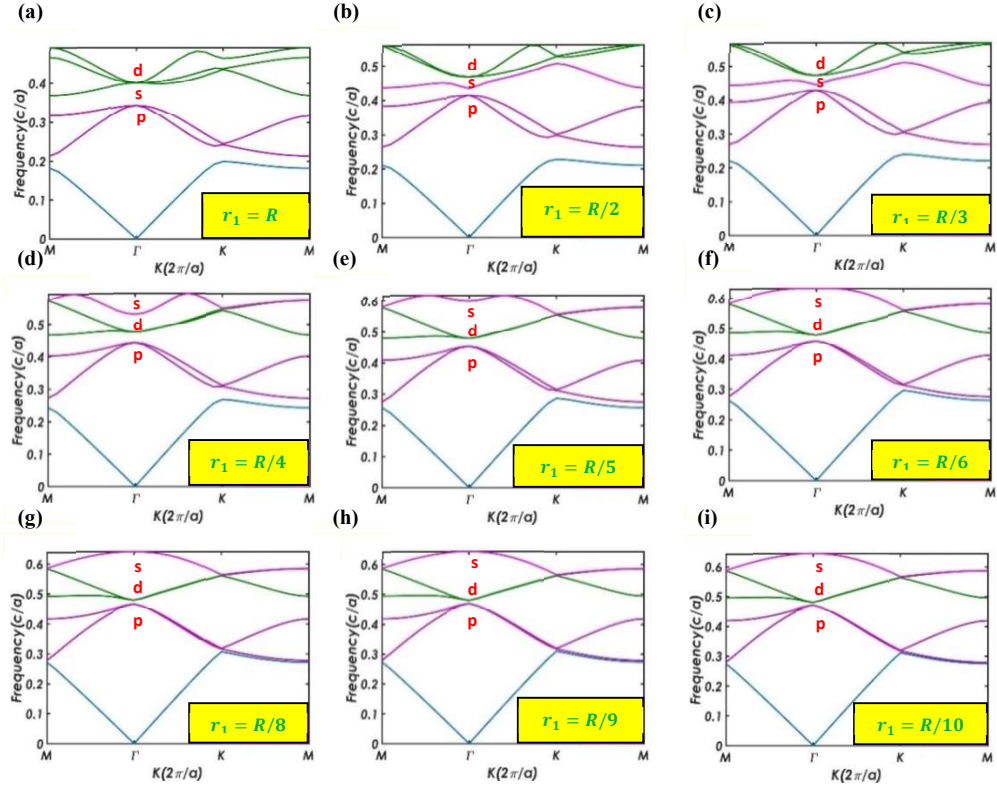

Fig. 3. The band diagrams of the modified honeycomb PCs for the radii values of (a) $r_1 = R$; (b) $r_1 = R/2$; (c) $r_1 = R/3$; (d) $r_1 = R/4$; (e) $r_1 = R/5$; (f) $r_1 = R/6$; (g) $r_1 = R/8$; (h) $r_1 = R/9$ and (i) $r_1 = R/10$.

Fig. 4(a) indicates the band diagram of this configuration with oblique boundary between the two kinds of PCs. The helical edge sates appear at the common band gap frequencies of these PCs. Figs. 4(b-d) show that two edge states A with the spin-up and B with the spin-down, have the same electric field distributions but possessing opposite Poynting vectors.

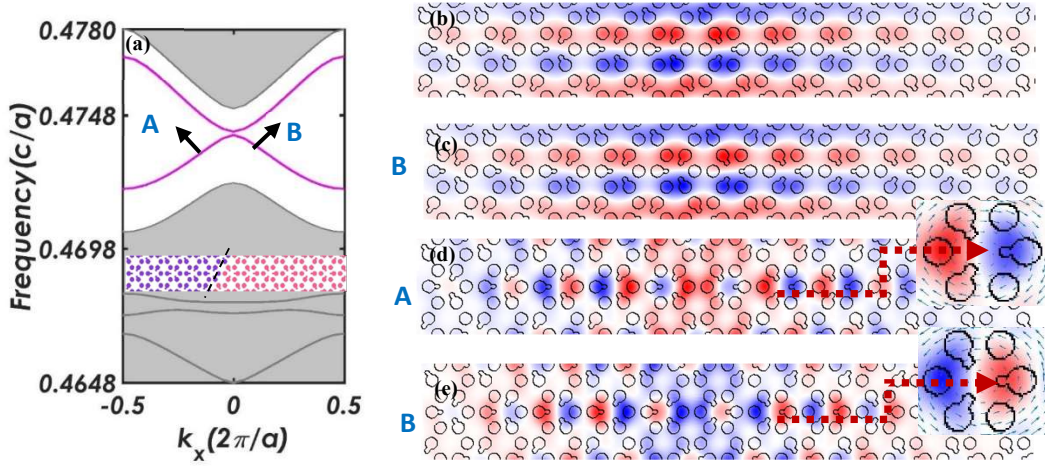

Fig. 4. (a) The band diagram of the left-right configuration of both modified honeycomb PCs with oblique boundary, its supercell is composed of $13 \times 2$ unit cells is in the inset, electric field outlines of the (b) spin-up, point A and (c) spin-down, point B, Poynting vectors of the (d) spin-up, point A and (e) spin-down, point B.

## 3. Photonic topological insulators

To study the one-way propagation of the edge states, we provide an up-down configuration of two PCs with perfectly matched layer (PML) as boundary conditions. The simulation is performed through the *Lumerical FDTD* solution [47]. As seen in Fig. 5(a), the light propagates along the interface without any noticeable wave scattering. We apply point like source at the interface to simulate the electric field propagation along the surface (yellow star). Indeed, two vertical magnetic dipoles with $90^0$ phase differences between them ($H_x + iH_y$) are applied at the edge state frequencies to puce corresponding electric field $E_z$. Insertion of additional elements inside the unit cell of honeycomb PC allows engineering of the topologically protected edge states via modifying degeneracies and photonic band gap characteristics. Besides, closing/opening the Dirac point and inversion of the profile of the electric field are main characteristic of the topological conversion. These occur via breaking the symmetry inversion of the honeycomb PC and make a new idea for studying the one-way propagation of the light along the interface of the ordinary topological PCs. To this aim, we propose a left-right geometry with the supercell consists of $13 \times 2$ unit cells of both modified honeycomb PCs to create a photonic topological insulator (PTI) with ordinary topological PCs acting as the bulk and edge states, respectively. Topological edge states can propagate against the defects, disorders and even local cavities without experiencing much scattering loss. To confirm this feature, we prepare a Z-shape interface and cavity configuration as seen in the Figs. 5(b)-5(c). Here, the cavity was made by removing an unit cell of PCs, types A and B, at the interface. In addition, the scattering-immune propagation is shown in two types of plasmonic defects [48]. The first defect is designed when one of the dielectric rods with $r = R = 0.1111\ \mu m$ is replaced with the silver (Ag) one as seen in Fig. 5(d) and in the second one, an Aluminum rectangular block with the length of $1.5\ \mu m$ and width of $50\ nm$ is embedded at $9\ \mu m$ from the source (Fig. 5(e)). We applied Palik [49] and Johnson and Christy models [50] for Ag and Al metals, respectively. Fig. 5(f) shows the comparison between the intensities versus frequency for various types of cavity and defects. The decreasing of the maximum intensity is seen clearly in propagating of the edge states at the interfaces involving the Ag rod, Al block or cavity.

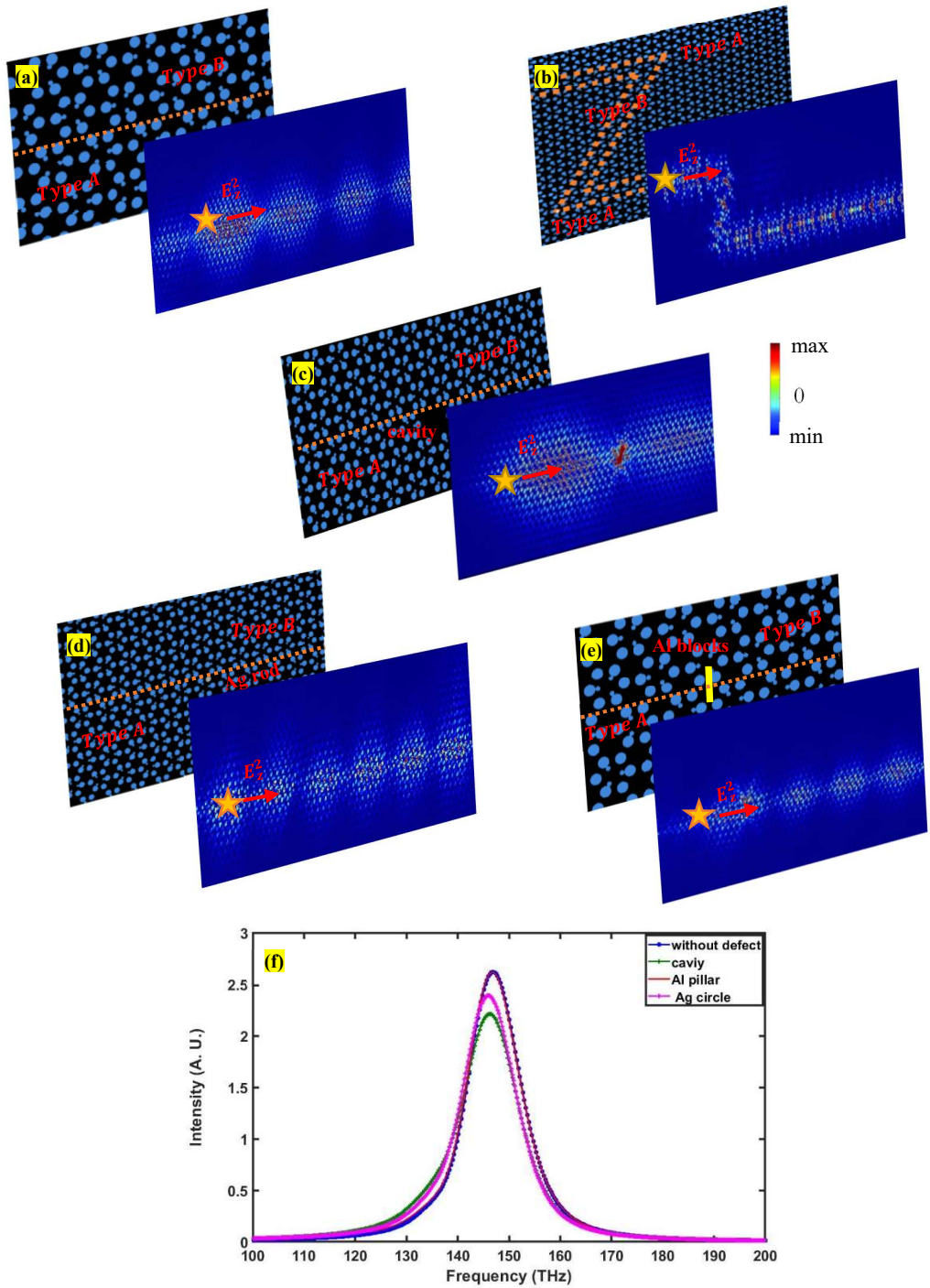

Fig. 5. (a) Unidirectional propagation of the circularly polarized source along the interface of two types of HPCs; (b)-(e) helical edge state distributions at the interface, Z shape, against the cavity, Ag rod and Al blocks, respectively. Yellow star indicates the point-like source at the interface; (f) the intensity of the edge states versus frequency for different cases: without defect, cavity, Al and Ag defects.

Here we emphasize that, we have used two different software packages, Mit Photonics Bands (MPB) and Lumerical FDTD solution for our calculations. **MPB uses a frequency-domain method** for the bandstructure calculation. In contrast, Lumerical FDTD (finite difference time domain) uses the time-domain techniques for the calculation of transmission, intensity. Therefore, because of using the different approaches, the normalized frequency of the edge states $\approx 0.47\ (c/a)$ is extracted from the MPB may be changed slightly in FDTD calculations using the real frequencies in the THz domain which results in appearing the error in using the exact frequency of the edge states. We also study the tunable intensity of the various defects in Fig. 6. To this end, we modify the radius of one of the rods (belongs to the type A) embedded at the $x = 11.6667\ \mu m, y = -0.144379$ from the source, in the all-dielectric topological insulator already shown in Fig. 5(a). As it is seen in Fig. 6(a), the intensity of the edge states can be increased by increasing the radius of the rod. The transmitted intensity can be enhanced via altering defect parameters. The maximum intensity is related to the $r = 0.18\ \mu m$ and increased up to the 130% of the reference intensity for $r = 0.1111\ \mu m$. Moreover, we change the geometry of the mentioned rod ($r = 0.1111\ \mu m$ at the $x = 11.6667\ \mu m, y = -0.144379$ from the source). So, its position may be perturbed in two ways: first, by moving the rod along the $+y$ direction and second, by moving along the vector $\vec{a}_2$ ($60^0$ degree rotation around the x direction). As seen in Figs. 6(b)-(c), the intensity of the edge states will decrease to 82.08% and 64.64%, respectively.

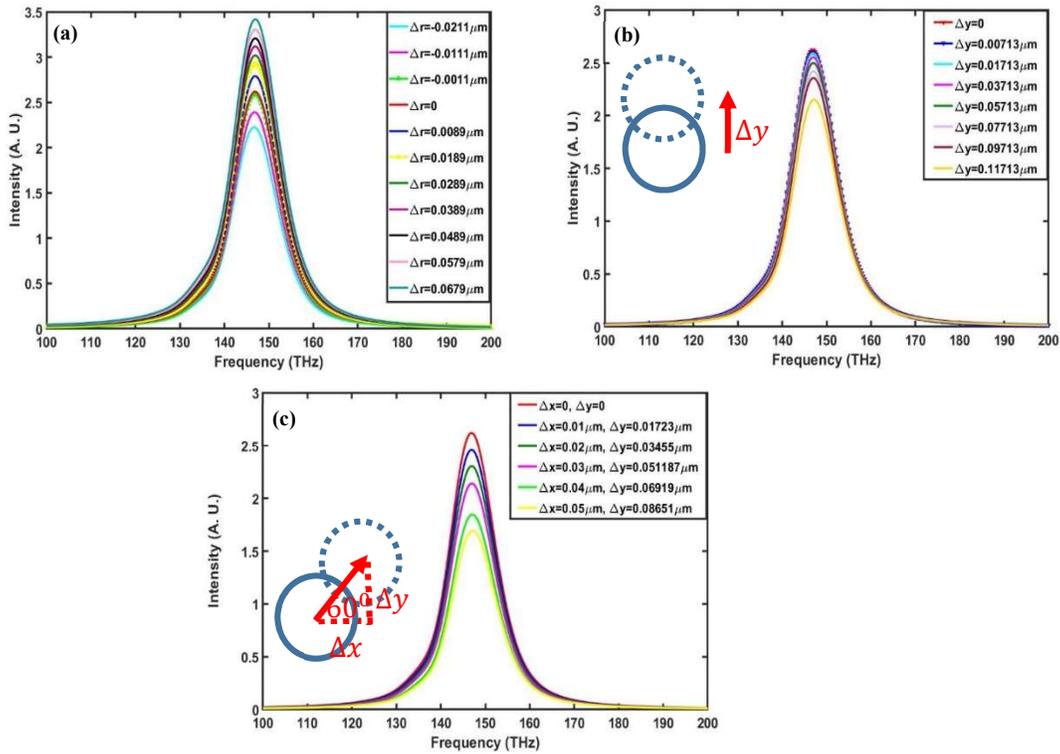

Fig. 6. The tunable intensity versus frequency for the case of all dielectric defect. (a) by increasing the radius, the intensity will be increase up to 130.62% of the reference number. (b)-(c) by increasing the distance of the defected rod along the directions, y and vector $\vec{a}_2$, the intensity will decrease to 82.08% and 64.64%, respectively.

In Fig.7 (a), again we study the intensity tuning in terms of defects, considering the Al blocks shown in Fig. 5(e) with the length of $0.2\ \mu m$ and the width in the range of $1-25\ nm$. Similar to the previous case, the intensity of the edge states, decrease by increasing the width of Al rectangular shape. Fig. 7 (b), indicates decreasing the normalized intensity by increasing the width of the Al blocks. Finally, we design two kinds of topological resonators in a rhombic shape. As seen in Fig. 8(a), the emitted light from the source channel, localizes around the resonator and cannot continue towards the right side of the interface. However, for the other modified resonator configuration, the light couples with the rhombic resonator, then propagates to the left side of the second interface as shown in Fig. 8(b). Besides, due to the topological robustness, there is not seen any back-scattering at the interfaces and the sharp corners of the rhombic do not deteriorate efficient light transmission towards the drop channel.

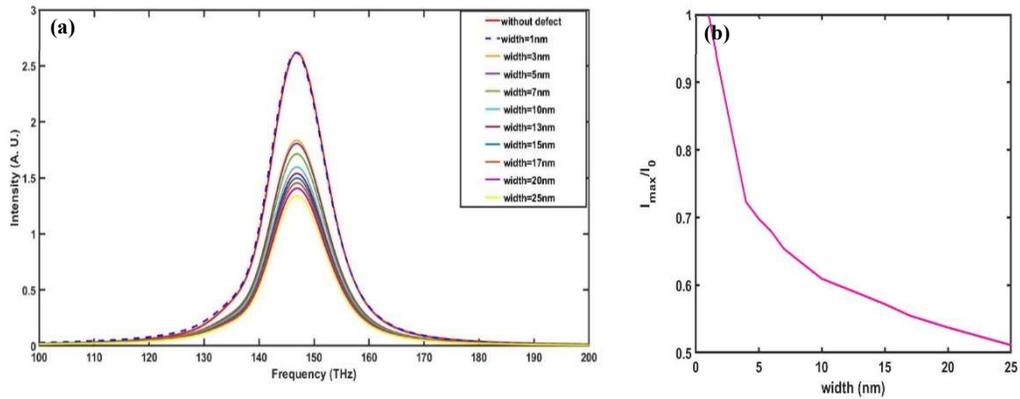

Fig. 7. (a) Variation of the intensity against the Al rectangular blocks defect; (b) decreasing the normalized by increasing the width of Al-blocks.

The intensities of electric field of the edge states versus frequency of the positions 1, 2 and 3 is shown in Fig. 8(c). As expected, the intensity of the edge states at the positions 1 and 2 are negligible in comparison with the intensity at the position 3. A little propagation observed at the right side (position 1), is due to the unidirectional propagation of the topological edge states at the interface. The frequency of the maximum intensity belongs the topological bandgap frequencies.

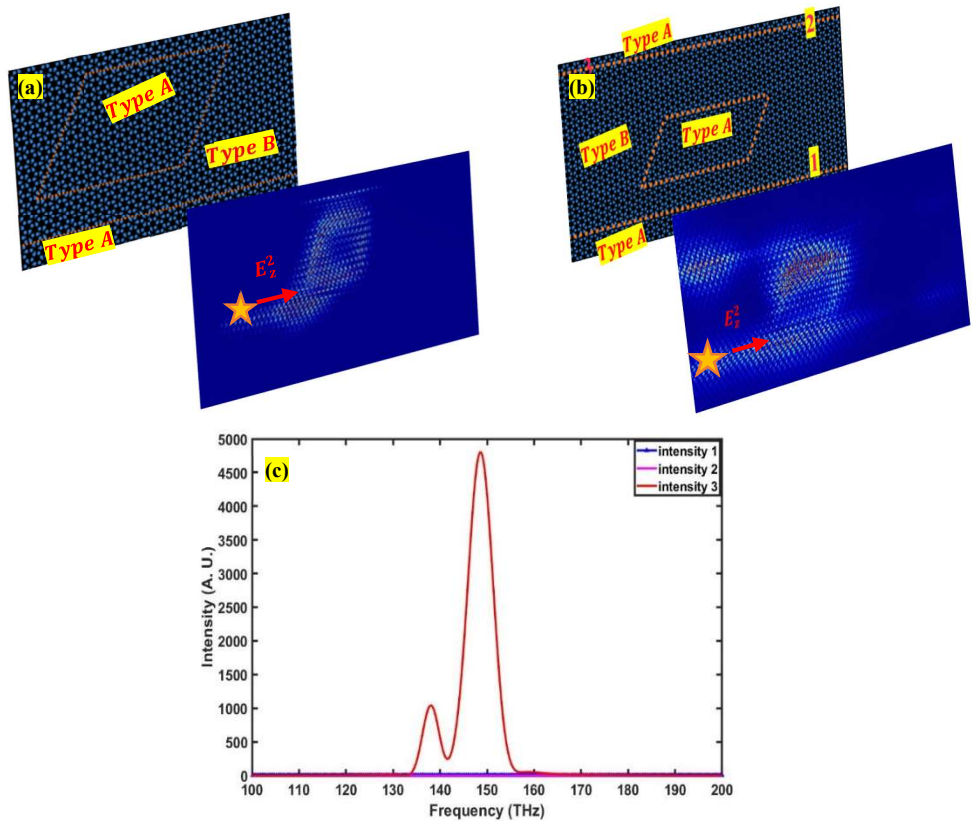

Fig. 8. The localized electric field intensity in the rhombic resonator and robustness of the topological edge states in the sharp corners of the rhombic. The dashed orange line shows the interface of the two types of PCs, A and B; (a) with one topological interface; (b) with two interfaces and (c) the intensity versus frequency at the positions 1, 2 and 3.

## 4. Conclusion

The topological behavior of the perturbed honeycomb PC has been studied by introducing three small dielectric rods in the neighborhood of the main ones in each unit cell of the structure. The double Dirac cone which emerges through the four-fold degeneracies at the $\Gamma$ point has been modified and the two-fold degeneracies of the bands are appeared. By studying the dispersion behavior of the photonic structure, we found that the big bandgap can be observed by choosing the radii of the small rods as $r_1 = R/7$. Closing and opening of the Dirac points which is the main characteristic of the topological transitions provide the useful platform for studying the intriguing phenomena of light like back-scattering immune transmission of the edge states along the different topological interfaces. We have shown efficient and various light propagation examples as well as tunable field intensity through disorders, defects and cavities. For example, three defects have been introduced in the structure: First: one of the dielectric rods with $r = 0.1111\ \mu m$ is replaced with the silver (Ag). Second: Aluminum rectangular blocks with the length of $1.5\ \mu m$ and width of $50\ nm$ is embedded at $9\ \mu m$ from the source. Third: Removing one of the rods and creating the cavity. Studying the intensities of the topological edge states in the mentioned defects and cavity, shows the decreasing of the intensity of the edge states through the interfaces involving the Ag rod, Al blocks or cavity. We also studied the tunable intensity of the edge states in the various defect structures. For example by increasing the radius of one of the rods (belongs to the type A) embedded at the $x = 11.6667\ \mu m, y = -0.144379$ from the source, we found that the intensity of the edge states can be enhanced. The maximum intensity is related to the $r = 0.18\ \mu m$ and increased up to the 130% of the reference intensity related to the $r = 0.1111\ \mu m$. Additionally, we modified the geometry of the mentioned rods with $r = 0.1111\ \mu m$ at the $x = 11.6667\ \mu m, y = -0.144379$ from the source. First, by moving the rods along the +y direction and second, by moving it along the vector $\vec{a}_2$ ($60^0$ degree rotation around the x direction). As a results, the intensity of the edge states will be decreased to 82.08% and 64.64%, respectively.

## Acknowledgments

H. K. acknowledges partial support of the Turkish Academy of Sciences.

## Disclosures

The authors declare no conflicts of interest.

## References

1. M. Z. Hasan and C. L. Kane, "Topological insulators," Rev. Mod. Phys. 82, 4 (2010).
2. X. L. Qi and S. C. Zhang, "Topological insulators and superconductors," Rev. Mod. Phys. 83, 4 (2011).
3. A. Bansil, H. Lin, and T. Das, "Colloquium: topological band theory," Rev. Mod. Phys. 88, 2 (2016).
4. K. V. Klitzing, G. Dorda, and M. Pepper, "New method for high-accuracy determination of the fine-structure constant based on quantized Hall resistance," Phys. Rev. Lett. 45, 6 (1980).
5. C. X. Liu, S. C. Zhang, and X. L. Qi, "The quantum anomalous Hall effect," Annu. Rev. Condens. Matter Phys. 7, 301 (2016).
6. C. L. Kane and E. J. Mele, "Quantum spin Hall effect in Graphene," Phys. Rev. Lett. 95, 22 (2005).
7. B. Y. Xie, H. F. Wang, X. Y. Zhu, M. H. Lu, Z. D. Wang and Y. F Chen, "Feature issue introduction: topological photonics and materials," Opt. Express 26, 19 (2018).
8. F. D. Haldane and S. Raghu, "Possible realization of directional optical waveguides in photonic crystals with broken Time-Reversal symmetry," Phys. Rev. Lett. 100, 013904 (2008).
9. Z. Wang, Y. D. Chong, J. D. Joannopoulos, and M. Soljacic, "Reflection-free one-way edge modes in a gyromagnetic photonic crystal," Phys. Rev. Lett. 100, 013905 (2008).
10. Z. Wang, Y. Chong, J. D. Joannopoulos, and M. Soljacic, "Observation of unidirectional backscattering-immune topological electromagnetic states," Nature (London) 461, 772 (2009).
11. K. Fang, Z. Yu, and S. Fan, "Realizing effective magnetic field for photons by controlling the phase of dynamic modulation," Nat. Photonics 6, 782 (2012).
12. D. L. Sounas, C. Caloz, and A. Alu, "Magnetic-free non-reciprocity based on staggered commutation," Nat. Commun. 4, 2407 (2013).

13. N. A. Estep, D. L. Sounas, J. Soric, and A. Alù, "Nonreciprocity and magnetic-free isolation based on optomechanical interactions," Nat. Phys. 10, 923 (2014).
14. M. A. Bandres, M. C. Rechtsman, and M. Segev, "Topological photonic quasicrystals: Fractal topological spectrum and protected transport," Phys. Rev. X 6, 011016 (2016).
15. M. C. Rechtsman, J. M. Zeuner, Y. Plotnik, Y. Lumer, D. Podolsky, F. Dreisow, S. Nolte, M. Segev, and A. Szameit, "Photonic floquet topological insulators," Nature (London) 496, 196 (2013).
16. D. Leykam and Y. D. Chong, "Edge solitons in nonlinear-photonic topological insulators," Phys. Rev. Lett. 117, 143901 (2016).
17. D. Leykam, M. C. Rechtsman, and Y. D. Chong, "Anomalous topological phases and unpaired Dirac cones in photonic floquet topological insulators," Phys. Rev. Lett. 117, 013902 (2016).
18. M. Hafezi, E. A. Demler, M. D. Lukin, and J. M. Taylor, "Robust optical delay lines with topological protection," Nat. Phys. 7, 907 (2011).
19. R. O. Umucalilar and I. Carusotto, "Artificial gauge field for photons in coupled cavity arrays," Phys. Rev. A 84, 043804 (2011).
20. M. Hafezi, S. Mittal, J. Fan, A. Migdall, and J. M. Taylor, "Imaging topological edge states in silicon photonics," Nat. Photonics 7, 1001 (2013).
21. M. C. Rechtsman, Y. Plotnik, J. M. Zeuner, D. Song, Z. Chen, A. Szameit, and M. Segev, "Topological creation and destruction of edge states in photonic Graphene," Phys. Rev. Lett. 111, 103901 (2013).
22. A. B. Khanikaev, S. H. Mousavi, W. K. Tse, M. Kargarian, A.H. MacDonald, and G. Shvets, "Photonic topological insulators, " Nat. Mater. 12, 233 (2013).
23. W. J. Chen, S. J. Jiang, X. D. Chen, B. Zhu, L. Zhou, J.W. Dong, and C. T. Chan, "Experimental realization of photonic topological insulator in a uniaxial metacrystal waveguide," Nat. Commun. 5, 5782 (2014).
24. H. Fan, B. Xia, L. Tong, S. Zheng, and D. Yu, "Elastic higher-order topological insulator with topologically protected corner states," Phys. Rev. Lett. 122, 204301 (2019).
25. A. B. Khanikaev, S. H. Mousavi, W. K. Tse, M. Kargarian, A. H. MacDonald & G. Shvets, "Photonic topological insulators," Nat. Mater. 12, 233 (2013).
26. T. Ma and G. Shvets, "All-Si valley-Hall photonic topological insulator," New J. Phys. 18, 025012 (2016).
27. L. H. Wu and X. Hu, "Scheme for achieving a topological photonic crystal by using dielectric material," Phys. Rev. Lett. 114, 223901 (2015).
28. C. He, X. C. Sun, X. P. Liu, M. H. Lu, Y. Chen, L. Feng, and Y. F. Chen, "Photonic topological insulator with broken time-reversal symmetry," Proc. Natl. Acad. Sci. USA 113, 4924 (2016).
29. L. Lu, J. D. Joannopoulos, and M. Soljačić, "Topological Photonics," Nat. Photon 8, 821 (2014).
30. X. C. Sun, C. He, X. P. Liu, M. H. Lu, S. N. Zhu, and Y. F. Chen, "Photonics meets topology," Prog. Quantum Electron 55, 52 (2017).
31. K. Sakoda, "Dirac cone in two and three-dimensional metamaterials," Opt. Express 20, 3898 (2012).
32. S. Raghu and F. D. M. Haldane, "Analogs of quantum Hall effect edge states in photonic crystals," Phys. Rev. A78, 033834 (2018).
33. L. Xu, H. X. Wang, Y. D. Xu, H. Y. Chen, and J. H. Jiang, "Accidental degeneracy in photonic bands and topological phase transitions in two-dimensional core-shell dielectric photonic crystals," Opt. Express 24, 18059 (2016).
34. H. X. Wang, L. Xu, H. Y. Chen, and J. H. Jiang, "Three-dimensional photonic Dirac points stabilized by point group symmetry," Phys. Rev. B 93, 235155 (2016).
35. H. X. Wang, Y. Chen, Z. H. Hang, H. Y. Kee, and J. H. Jiang, "Type-II. Dirac photons," npj Quantum Mater 2, 54 (2017).
36. S. Barik, H. Miyake, W. DeGottardi, E. Waks, and M. Hafezi, "Two-dimensionally confined topological edge states in photonic crystals," New J. Phys. 18, 113013 (2016).
37. Y. Yang, Y. F. Xu, T. Xu, H. X. Wang, J. H. Jiang, X. Hu, and Z. H. Hang, arXiv:1610.07780.
38. J. Noh, W. A. Benalcazar, S. Huang, M. J. Collins, K. Chen, T. L. Hughes, and M. C. Rechtsman, "Topological protection of photonic mid-gap defect modes," Nat. Photonics 12, 408 (2018).
39. X. Zhu, H. X. Wang, C. Xu, Y. Lai, J. H. Jiang and S. John, "Topological transitions in continuously deformed photonic crystals," J. Phys. Rev. B 97, 085148 (2018).
40. J. Hajivandi and H. Kurt, "Robust transport of the edge modes along the photonic topological interfaces of different configurations", arXiv:2005.08775 [physics.app-ph] (2020).
41. J. Hajivandi and H. Kurt, "Topological phase transition of the centered rectangular photonic lattice", arXiv:2005.11916 [physics.app-ph] (2020).

42. J. Hajivandi and H. Kurt, "Topological photonic states and directional emission of the light exiting from the photonic topological structure composed of two dimensional honeycomb photonic crystals with different point group symmetries", arXiv:2002.11979 [physics.app-ph] (2020).
43. J. Hajivandi, E. Kaya, G. Edwards and H. Kurt, "Simulating topological robustness of Fano resonance in rotated Honeycomb photonic crystals", arXiv:2003.07253 [physics.app-ph] (2020).
44. S. Barik, H. Miyake, W. DeGottardi, E. Waks and M. Hafezi, "Two-dimensionally confined topological edge states in photonic crystals," New J. Phys. 18, 113013 (2016).
45. S. Barik et.al., "A topological quantum optics interface," Science 359, 666 (2018).
46. S. G. Johnson and J.D. Joannopoulos, "Block-iterative frequency-domain methods for Maxwell's equations in a planewave," basis. Opt. Express 8, 173 (2001).
47. Lumerical Inc. http://www.lumerical.com/tcad-pucts/fdtd/
48. Z. Song, H. J. Liu, N. Huang and Z. Wang, "Prediction of two-dimensional organic topological insulator in metal-DCB lattices," Appl. Optics 57, 29 (2018).
49. E. D. Palik. (1985). *Handbook of optical constants of solids*. Orlando: Academic Press.
50. P. B. Johnson and R. W. Christy, "Optical Constants of the Noble Metals," Phys. Rev. B 6, 4370 (1972).